\documentclass[
prl,
twocolumn,
amsmath,
superscriptaddress,
notitlepage
,longbibliography
]{revtex4-2}

\usepackage{graphicx,amssymb,amsmath,amsfonts}
\graphicspath{ {./figures/} }
\usepackage{xcolor}
\usepackage{mhchem}
\mhchemoptions{mathfontcommand=\mathit}

\def\papertitle{Crowding-regulated binding of divalent biomolecules}
\def\papertitle{Crowding-Regulated Binding of Divalent Biomolecules}

\usepackage[draft,inline,nomargin,final]{fixme}

\newcommand{\nn}{\nonumber\\}

\fxsetup{theme=color,mode=multiuser}
\FXRegisterAuthor{ts}{ats}{\color{orange}TS}
\FXRegisterAuthor{sk}{ask}{\color{red}SK}
\FXRegisterAuthor{mj}{amj}{\color{blue}MJ}
\FXRegisterAuthor{ac}{aac}{\color{green}AC}

\usepackage{xspace}
\newcommand{\latin}[1]{{\it #1}}
\newcommand{\ie}{\latin{i.e.}\@\xspace}
\newcommand{\eg}{\latin{e.g.}\@\xspace}
\newcommand{\cf}{\latin{cf.}\@\xspace}
\newcommand{\etc}{\latin{etc}\@\xspace}

\newcommand{\vs}{\latin{vs.}\@\xspace}

\newcommand{\ssect}[1]{Sect.~\ref{sm:#1} in \footnotemark[1]}
\newcommand{\scref}[1]{\cref{#1} in \footnotemark[1]}

\newcommand{\focc}{\phi}

\newcommand{\cb}{{\left[ \ce{B} \right]}}

\newcommand{\caa}{{\left[ \ce{AA} \right]}}
\newcommand{\caab}{{\left[ \ce{AA*B} \right]}}
\newcommand{\caabb}{{\left[ \ce{AA*B_2} \right]}}

\newcommand{\gb}{\gamma_{\ce{B}}}

\newcommand{\gaa}{\gamma_{\ce{AA}}}
\newcommand{\gaab}{\gamma_{\ce{AA*B}}}
\newcommand{\gaabb}{\gamma_{\ce{AA*B_2}}}

\newcommand{\raa}{R_{\ce{AA}}}
\newcommand{\raab}{R_{\ce{AA*B}}}
\newcommand{\raabb}{R_{\ce{AA*B_2}}}

\newcommand{\Nc}{N_\mathrm{c}}
\newcommand{\Rc}{R_\mathrm{c}}

\newcommand{\ftitle}[1]{{\bf #1}}
\newcommand{\fsub}[1]{({\bf #1})}

\newcommand{\mycite}[1]{ref.~\cite{#1}}

\usepackage{siunitx}
\DeclareSIUnit[number-unit-product = {}]
  \poise{P}
\DeclareSIUnit\molar{M}
\DeclareSIUnit[number-unit-product = {}]
  \calorie{cal}

\usepackage{xr}
\usepackage{hyperref}
\usepackage[capitalize]{cleveref}

\begin{document}

\title{\papertitle}

\author{Tomasz Sk\'ora}
\affiliation{Institute of Physical Chemistry, Polish Academy of Sciences, 01-224~Warsaw, Poland}
\affiliation{Scientific Computing and Imaging Institute, University of Utah, 84112~Salt Lake City, Utah, United States}

\author{Mathijs Janssen}
\affiliation{Department of Mathematics, Mechanics Division, University of Oslo, N-0851 Oslo, Norway}
\affiliation{Centre for Cancer Cell Reprogramming, Faculty of Medicine, University of Oslo, Montebello, N-0379 Oslo, Norway}
\affiliation{Norwegian University of Life Sciences, Faculty of Science and Technology, Pb 5003, 1433 Ås, Norway}

\author{Andreas Carlson}
\affiliation{Department of Mathematics, Mechanics Division, University of Oslo, N-0851 Oslo, Norway}

\author{Svyatoslav Kondrat}
\affiliation{Institute of Physical Chemistry, Polish Academy of Sciences, 01-224~Warsaw, Poland}
\affiliation{Institute for Computational Physics, University of Stuttgart, Stuttgart, Germany}

\date{\today}

\begin{abstract}

Macromolecular crowding affects biophysical processes as diverse as diffusion, gene expression, cell growth, and senescence. 
Yet, there is no comprehensive understanding of how crowding affects reactions, particularly multivalent binding.
Herein, we use scaled particle theory and develop a molecular simulation method to investigate the binding of monovalent to divalent biomolecules.
We find that crowding can increase or reduce cooperativity---the extent to which the binding of a second molecule is enhanced after binding a first molecule---by orders of magnitude, depending on the sizes of the involved molecular complexes.
Cooperativity generally increases when a divalent molecule swells and then shrinks upon binding two ligands.
Our calculations also reveal that, in some cases, crowding enables binding that does not occur otherwise.
As an immunological example, we consider Immunoglobulin G-antigen binding and show that crowding enhances its cooperativity in bulk but reduces it when an Immunoglobulin G binds antigens on a surface.

\end{abstract}

\maketitle

\footnotetext[1]{See Supplementary Material, which includes \mycite{Janssen2021a, Diestler2010, Gibbons1969,Minton1998, Ando2010, Huber2019, Slyk2022b, Bongini2004, Saphire2001, Berman2000, Sehnal2021, Chen2002, Rubinstein2003}}

\paragraph{Introduction.}
The intracellular space of living cells is crowded by biomacromolecules---in \emph{E. coli}, for instance, up to $44\%$ \cite{Luby-Phelps2013,Minton2015,Zimmerman1991}.
These macromolecular ``crowders" can affect biophysical processes through the volume they exclude:
\emph{crowding} \cite{Zimmerman1993,Minton1998,Ellis2001,Feig2017} can hinder diffusion \cite{Dix2008,Junker2019,Skora2020}, promote association reactions \cite{Minton2005a,Schreck2020,Heo2022},\phantomsection{\label{ref:schreck}}~shift equilibria in the direction of smaller and more spherical conformations of reacting species \cite{Cheung2005,Gasic2019}, enhance or reduce enzyme-catalysed reactions \cite{Norris2011, Pastor2014, Maximova2019, Skora2021}, \etc.
As a result, crowding affects physiological processes like senescence \cite{Minton2020}, gene regulation \cite{Tabaka2014}, and cell growth \cite{Alric2022}.

Biochemical reactions often involve macromolecules with more than one binding site \cite{Hunter2009,Janssen2021a,Ercolani2011,Perelson1980}.
Examples of \emph{multivalent binding} include hemoglobin binding four oxygen molecules \cite{Ackers1992,Eaton1999}, antibodies binding two antigens \cite{Yang2017}, the condensation of multivalent intrinsically-disordered proteins \cite{Li2012,Borcherds2021}, and multivalent binding in synthetic systems \cite{Shuker1996,Mack2011}.
Multivalent binding can be \emph{cooperative}, meaning that the association constant of each further binding step is larger than that of the previous step \cite{Whitty2008,Hunter2009}.
Cooperative systems show ``on-off behavior" where multivalent molecules change from being unbound to being fully bound upon slight changes in, \eg, ligand concentration or temperature; noncooperative systems behave more gradually.
\sknote{We can remove this sentence if needed space}Correspondingly, the concentration of partly-bound  multivalent particles is smaller for more cooperative binding reactions.
One speaks of \emph{allosteric} cooperativity when the binding interactions are intermolecular, as is the case, for instance, with oxygen-hemoglobin binding.
Conversely, the cooperativity of intramolecular interactions is called \emph{chelate} \cite{Ercolani2011}.

Despite the biological importance of molecules with multiple binding sites, how crowding affects multivalent binding is yet to be investigated theoretically (see, however, \mycite{Lei2015}).
Herein, we use scaled particle theory (SPT) and develop a general molecular simulation method to study how physiologically relevant crowding conditions affect divalent binding.
We find that crowding can enhance or reduce cooperativity depending on the size differences between reactants and products, in some cases enabling divalent binding per se. 
With molecular simulations of a coarse-grained Immunoglobulin G (IgG) model, we show that IgG-antigen binding cooperativity depends sensitively on whether it occurs in bulk or on a surface.

\begin{figure*}[]
	\includegraphics[width=\textwidth]{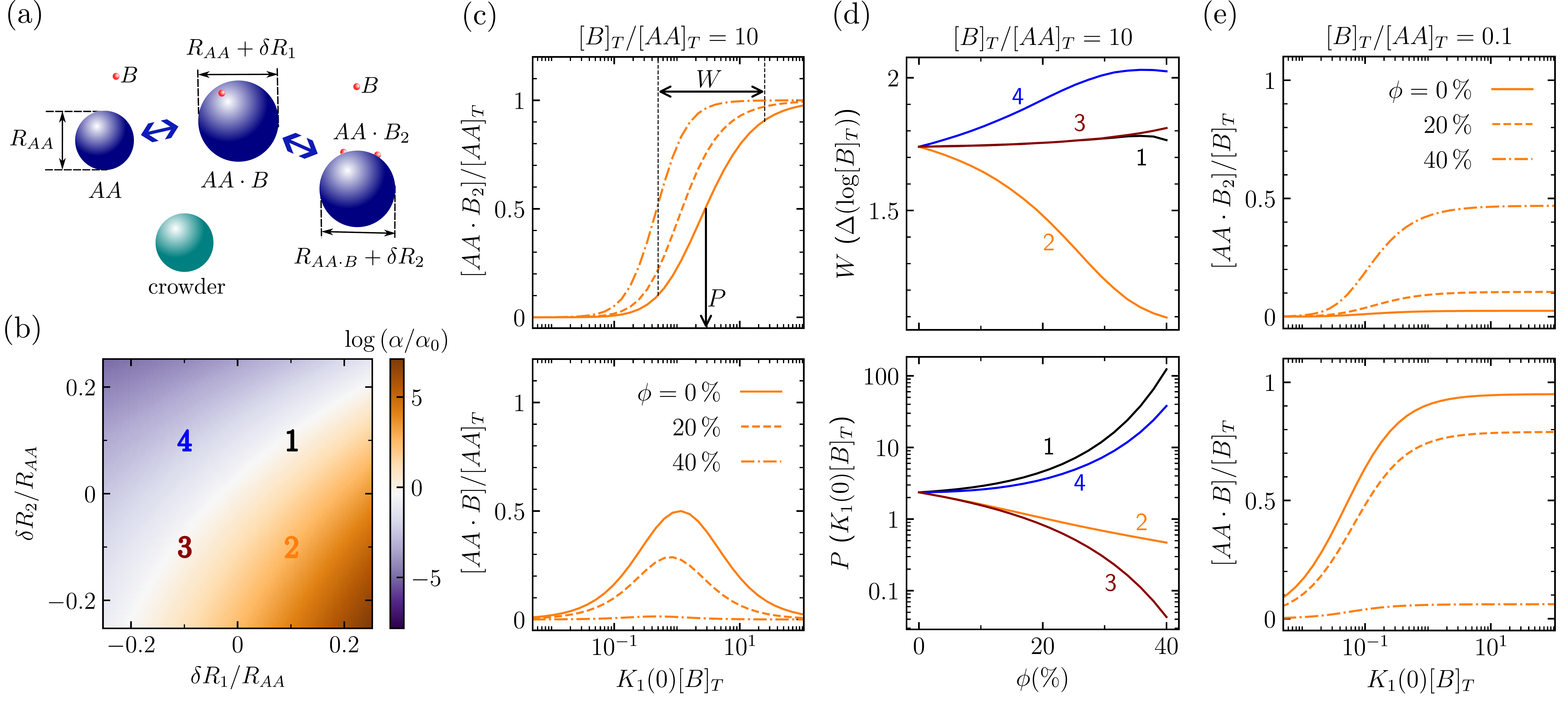}
    \centering
    \caption{
	    \ftitle{Divalent binding with crowding dependence through SPT theory.} 
	    \fsub{a} Schematic of a divalent molecule \ce{AA} that binds two monovalent molecules \ce{B}, indicating the sizes of the different macromolecular complexes.
	    \fsub{b} Heat map of $\log (\alpha / \alpha_0)$ plotted in the plane of $\delta R_1 / \raa$ and $\delta R_2 / \raa$ for an occupied volume fraction $\focc = \SI{40}{\percent}$, \ce{AA} size $\raa=\SI{6}{\nano\meter}$, and crowder size $\Rc = \SI{5.1}{\nano\meter}$ (corresponding to Ficoll 70, a typical synthetic crowder). 
	    The coloured numbers (1-4) indicate the values of $\delta R_1=\pm0.1\raa$ and $\delta R_2=\pm0.1\raa$ used in the other panels.
	    \fsub{c} Dependence of $\caabb$ and $\caab$ on the concentration of B molecules. 
	    The concentration $\cb_T/\caa_T=10$ and $\delta R_1 = -\delta R_2 = 0.1 \raa$ (point 2 in panel (b)).
	    \fsub{d} The width and position of the $\caabb$ profile (\cf panel (c)) for $\cb_T/\caa_T = 10$ and for four combinations of $\delta R_1=\pm0.1$ and $\delta R_2\pm0.1$ corresponding to the points in panel (b).
	    \fsub{e}  Dependence of $\caabb$ and $\caab$ on the concentration of B molecules for $\cb_T/\caa_T=0.1$ and $\delta R_1 = -\delta R_2 = 0.1 \raa$ (point 2 in panel (b)).
	}
\label{fig:allosteric:Minton}
\end{figure*}

\paragraph{Divalent binding.} 
We consider a divalent molecule \ce{AA} that reversibly binds two \ce{B} molecules (\cref{fig:allosteric:Minton}a),
\begin{subequations}
	\label{eq:reaction}
	\begin{align}
		\ce{AA + B &<=>[2$K_1$] AA*B} ,\\
		\ce{AA*B + B &<=>[\frac{1}{2}$K_2$] AA*B_2} ,
	\end{align}
\end{subequations}
where $K_1$ and $K_2$ are equilibrium constants of the individual binding events. 
We denote by $\caa$ and $\cb$ the equilibrium concentrations of unbound \ce{AA} and \ce{B} molecules and by $\caa_T=\caa+\caab+\caabb$ and  $\cb_T=\cb+\caab+2\caabb$ their total concentrations, where $\caab$ and $\caabb$ are the equilibrium concentrations of \ce{AA*B} and \ce{AA*B_2} complexes.\phantomsection{\label{ref:equilibrium}}~The reaction-rate equations associated with \cref{eq:reaction} are a set of four coupled ordinary differential equations for the time-dependent concentrations of the four molecular species.
At steady state, these equations can be reduced to a single cubic equation for $\caab$ (see \mycite{Janssen2021a}) or $\cb$ (see \mycite{Kaufman4157,reynolds1979}) in terms of $K_1$, $K_2$, $\caa_T$, and $\cb_T$. 
The solutions to these cubic equations then allow one to find the concentrations of the other species (\ssect{sec:conc}).

If the above reaction happens among chemically inert crowders, the association constants amount to
\begin{subequations}
	\label{eq:Kact}
	\begin{align}
		K_1 \left( \focc \right)& = \frac{1}{2}\frac{\caab}{\caa\cb} = K_1 \left( 0 \right) \frac{\gaa\gb}{\gaab} ,\\
		K_2 \left( \focc \right) &= 2\frac{\caabb}{\caab\cb} = K_2 \left( 0 \right) \frac{\gaab\gb}{\gaabb} ,
	\end{align}
\end{subequations}
where $\focc$ is the volume fraction occupied by crowders and $\gamma_i$ is the activity coefficient of species $i\in\{\ce{AA},\ce{B},\ce{AA*B},\ce{AA*B2}\}$. 
We use the allosteric cooperativity factor $\alpha=K_2/K_1$ \cite{Hunter2009} and
speak of cooperative binding when $\alpha>1$. 
With \cref{eq:Kact}, we find
\begin{equation}
	\label{eq:alpha}
	\frac{\alpha}{\alpha_0} = \frac{\gaab^2}{\gaa \gaabb},
\end{equation}
where $\alpha_0 = K_2(0)/K_1(0)$ is the cooperativity factor in the absence of crowders. 
If, instead of two monovalent \ce{B} molecules, \ce{AA} binds another divalent molecule \ce{BB}, the second binding step is intramolecular.
For this case, the chelate cooperativity factor is $\alpha=K_2/(2[\ce{BB}]K_1)$ \cite{Ercolani2011}, yielding the same expression for $\alpha/\alpha_0$ but with a different $\alpha_0$.
In the following, we consider $\caa_T$ and $\cb_T$ much smaller than the concentration of crowders and study how crowding affects cooperativity by computing activity coefficients as functions of $\focc$.

\paragraph{SPT results.}
As a starting point, we use SPT, which offers approximate analytical expressions for the activity coefficients of macromolecules and macromolecular complexes of convex shapes.
A protein-ligand complex can have a larger \cite{Mangel1990,Tsytlonok2020} or a smaller \cite{Pickover1979,Kumar1980,Newcomer1981,Dumas1983,Shaanan1983,Olah1993,Taylor1996,Sevvana2008,Salvi2017,Ghobadi2018} radius of gyration than the corresponding unbound protein.
Accordingly, in our model, we consider \ce{AA}, \ce{AA*B}, and \ce{AA*B2} complexes to be spheres of radii $\raa$, $\raab = \raa + \delta R_1$, and $\raabb = \raab + \delta R_2$, respectively (\cref{fig:allosteric:Minton}a), where $\delta R_1$ and $\delta R_2$ can be positive or negative.
Assuming also crowders to be spheres of equal radius $\Rc$, we can estimate the activity coefficients $\gamma_i$ with Minton's generalized SPT \cite{Minton1998} (\ssect{sec:SPT})
\begin{align}\label{eq:spt_transformed}
	\ln \gamma_i(\focc) &= \ln \left( 1 +g \right) 
		+ g \left( \frac{3R_i}{\Rc} + \frac{3R_i^2}{\Rc^2} 
		+ \frac{R_i^3}{\Rc^3} \right) \nn
		&\quad+ \frac{g^2}{2} \left( \frac{9R_i^2}{\Rc^2} + \frac{6R_i^3}{\Rc^3} \right) 	
		+ 3 g^3 \frac{R_i^3}{\Rc^3},
\end{align}
where $g(\focc) = \focc/(1 - \focc)$.
We combine \cref{eq:alpha,eq:spt_transformed} to draw a heat map (\cref{fig:allosteric:Minton}b) of $\log(\alpha/\alpha_0)$ (base 10) in the plane of $\delta R_1$ and $\delta R_2$ for $\focc=\SI{40}{\percent}$.
We see that $\ce{AA}$ particles that swell and then shrink upon binding two $\ce{B}$ particles (point 2) have an enhanced cooperativity. 
By contrast, $\ce{AA}$ particles that shrink and then swell (point 4) have a decrease in cooperativity. 
The cooperativity of particles that swell twice (point 1) or shrink twice (point 3) is hardly affected by crowders.
Intuitively, excluded-volume interactions with crowders become more important for larger molecules because they leave less space for the crowders. 
In the case of point 2, with intermediates \ce{AA*B} being larger than \ce{AA} and \ce{AA*B_2}, crowder-molecule interactions will suppress $\caab$, driving up $\caa$ and $\caabb$.
Hence, $K_1(\phi)$ decreases and $K_2(\phi)$ increases with $\phi$, explaining the $\log(\alpha/\alpha_0)>0$ value observed.
Similar reasoning explains the trends at points 1, 3, and 4.

Next, we calculate binding curves of $\caabb/\caa_T$ and $\caab/\caa_T$ \vs the scaled density $K_1\cb_T$ at fixed $\cb_T/\caa_T$.
These curves represent numerical solutions to the cubic equation mentioned below \cref{eq:reaction} (see \scref{sm:eq:cubic})---we again used Minton's theory to determine $K_{1}(\focc)$ and $K_{2}(\focc)$. 
\Cref{fig:allosteric:Minton}c concerns a case where \ce{B} molecules outnumber \ce{AA} molecules, $\cb_T/\caa_T=10$.
We also set $\delta R_1=0.1\raa$ and  $\delta R_2=-0.1\raa$, corresponding to point 2 in \cref{fig:allosteric:Minton}b. 
The $\caabb/\caa_T$ profile has a typical sigmoidal shape that shifts toward smaller $K_{1}(0)\cb$ with increasing $\focc$.
The sigmoid also becomes steeper with $\focc$, which is the on-off behavior typical of cooperative systems mentioned in the introduction.
Moreover, $\caab$ decreases with crowding---at the highest $\focc=\SI{40}{\percent}$ considered, there are practically no intermediates \ce{AA*B} at any concentration of monovalent molecules (bottom plot in \cref{fig:allosteric:Minton}c).

To characterise a $\caabb/\caa_T$ curve, we define the position $P$ of its midpoint as the $K_{1}(0)\cb$ value at which the derivative $d \caabb/d\cb_T$ is maximal.
We also define the profile width $W$ as the difference in the $K_{1}(0)\cb$ values at which the concentrations $\caabb/\caa_T$ are $10/11$ and $1/11$ of the maximum (\ie, $\caa_T$).
Note that $W$ is similar to the switching window of Hunter and Anderson \cite{Hunter2009}, which dealt with the total concentration of bound molecules (\ie, the sum of $\caabb$ and $\caab$) instead. 
In \cref{fig:allosteric:Minton}c (top panel), $W$ decreases with crowding, in line with the increased cooperativity at point 2 in \cref{fig:allosteric:Minton}b. 
The position $P$ decreases with increasing $\focc$
because crowders tend to promote binding: for given $K_1\cb_T$, a larger fraction of $\ce{AA}$ particles are in the fully bound state $\ce{AA*B2}$.

\Cref{fig:allosteric:Minton}d shows $P$ and $W$ \vs $\focc$ for the four combinations of $\delta R_1$ and $\delta R_2$ indicated with numbers 1 to 4 in panel (b). 
The panel for $W$ corroborates our panel (b) findings: $W$ varies inversely with the change $\alpha/\alpha_0$ in cooperativity, where molecules that swell and shrink have large $\alpha/\alpha_0$, hence their $W$ decreases. 
Molecules for which $\delta R_1= \delta R_2$ (points 1 and 3) showed no change in cooperativity in panel (b), and, likewise, their window $W$ \vs $\focc$ is roughly constant.
Interestingly, the positions $P$ of their binding curves show opposing trends. 
The conventional wisdom of crowders driving reactions towards the bound state (\ie decreasing $P$) does not hold for divalent molecules that continually swell upon binding and is, thus, not correct in general. This conclusion is in line with a recent study by \citeauthor{Schreck2020} \cite{Schreck2020} on protein folding and aggregation.

Lastly, \cref{fig:allosteric:Minton}e shows binding curves for the same settings as in panels (c) and (d) except $\cb_T/\caa_T=0.1$; hence, \ce{AA} now outnumber \ce{B} molecules. 
In absence of crowders and for large $K_{1}(0)\cb$, the system is saturated with intermediates \ce{AA*B}, while the concentration of \ce{AA*B_2} practically vanishes.
This situation changes when the occupied volume fraction  $\focc$ of crowders increases, especially for the physiologically-relevant value $\SI{40}{\percent}$.
In that case, the concentration of fully bound complexes \ce{AA*B2} reaches values close to the maximum $\caabb \approx 0.5 \cb_T$, while the concentration of intermediates almost vanishes.
This behavior is because crowders enhance cooperativity (for $\delta R_1 = -\delta R_2 = 0.1 \raa$ as considered here), and, in turn, cooperativity promotes fully-bound complexes. 
Hence, when \ce{AA} molecules outnumber \ce{B} molecules, crowding enables divalent binding.

\paragraph{Activity coefficients from molecular simulations.}
SPT is limited to convex shapes and breaks down for high volume fractions and small crowders.\phantomsection{\label{ref:mc}}~
Here, based on \mycite{Diestler2008}, we develop a method for numerically calculating activity coefficients in crowded media with Monte Carlo (MC) and Brownian dynamics (BD) simulations.
We consider a molecule of species $i$ to comprise $n_i$ beads. 
When such a molecule is among $\Nc$ crowders at positions $\{\boldsymbol{R}_k\}$, its activity coefficient reads \cite{Diestler2008,Diestler2010} (\ssect{sec:statmech})
\begin{subequations}\label{eq:gamma_complete}
\begin{align}
	\label{eq:gamma_i}
	\gamma_i^{-1} (\{\boldsymbol{R}_k\}) = Z_i(\{\boldsymbol{R}_k\}) Z_i^{-1}(\emptyset),
\end{align}
where $\{\boldsymbol{R}_k\}=\emptyset$ signifies the absence of crowders and 
\begin{align}\label{eq:gammastatmech}
	Z_i(\{\boldsymbol{R}_k\}) &= \int \prod_{j=1}^{n_i} d\boldsymbol{r}_{j} \exp \left[ -\beta \mathcal{U}_i \left( \left\{ \boldsymbol{r}_j \right\} \right) \right] \nn
	&\qquad\times\exp \left[ -\beta \sum_{j=1}^{n_i} \sum_{k=1}^{\Nc} U(\boldsymbol{r}_j - \boldsymbol{R}_k) \right] 
\end{align}
\end{subequations}
is the configurational partition function of the molecule, where $\mathcal{U}_i$ is the intramolecular potential energy of species $i$ and $U(\boldsymbol{r}_j - \boldsymbol{R}_k)$ is the interaction energy of crowder $k$ with bead $j$ at position  $\boldsymbol{r}_j$. 
To calculate the integral in \cref{eq:gamma_i} through the MC method, we reduce it to (\ssect{sec:MC})
\begin{align}
	\label{eq:mcinsertions}
	\gamma_i^{-1}(\{\boldsymbol{R}_k\})
	\approx \frac{1}{M} \sum_{\alpha=1}^{M} \exp \left[ -\beta \sum_{j=1}^{n_i} \sum_{k=1}^{\Nc} U \left(\boldsymbol{r}^{(\alpha)}_{j} - \boldsymbol{R}_k\right) \right] ,
\end{align}
where $\alpha$ in $\boldsymbol{r}^{(\alpha)}_{j}$ denotes an MC draw with conformations (\ie, $\boldsymbol{r}^{(\alpha)}_{j}$) obtained from BD simulations of single molecule $i$, and $M$ is the total number of draws (\ssect{sec:BD:1}). To obtain the activity coefficient in a crowded system, we average over the distribution of crowders, $\gamma_i^{-1} = \langle \gamma_i^{-1}(\{\boldsymbol{R}_k\}) \rangle_\mathrm{c}$, where we used BD simulation to obtain crowder configurations in the absence of molecule $i$ (\ssect{sec:BD:crowding}). 

While this approach is general, in all our calculations below, we consider only hard-sphere interactions $U=U_\mathrm{HS}$ (see \scref{sm:eq:harmonic}) between beads and crowders.
In this case, we have $\gamma_i^{-1}=V_{\mathrm{acc},i}/V$, with
$V$ the volume of the system and
$V_{\mathrm{acc},i}$ the average volume accessible to molecule $i$, that is, not excluded by crowders.
Thus, the activity of the molecular species $i$ is $a_i \equiv \gamma_i [i] = N_i /V_{\mathrm{acc},i}$, where $N_i$ is the number of molecules $i$. 

\begin{figure}[]
	\includegraphics[width=0.9\linewidth]{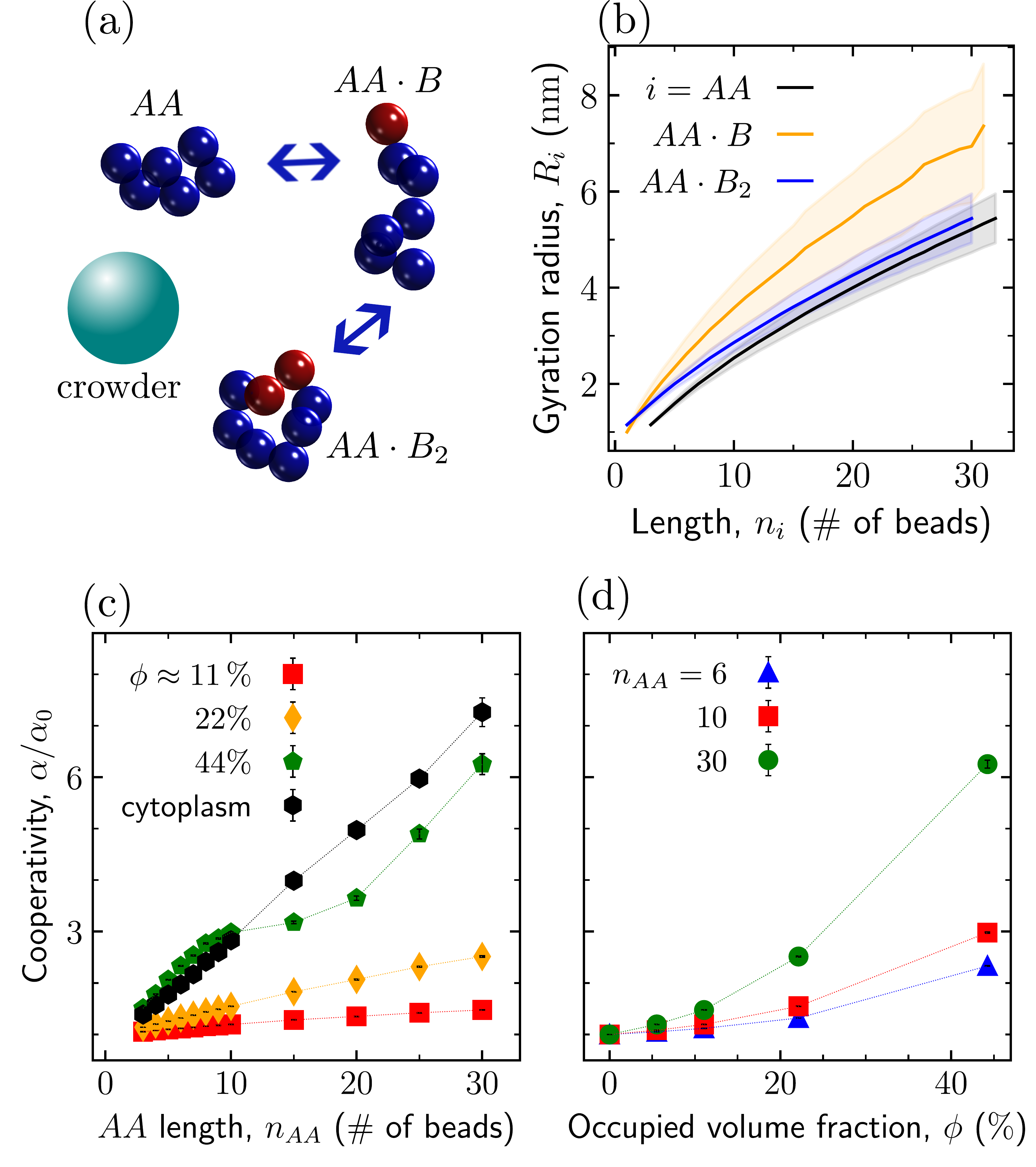}
    \centering
    \caption{
	    \ftitle{Cooperativity from simulations.} 
	    \fsub{a} Schematic of the binding of a monovalent molecule \ce{B} (red beads) to a divalent molecule \ce{AA} (composed of blue beads). 
	In the first step, a cyclic \ce{AA} opens to bind \ce{B}; in the second step, \ce{AA} binds a second \ce{B} molecule and closes to become a ring again.
	    \fsub{b} Gyration radii of various complexes.
	    \fsub{c,d} Cooperativity as a function of \fsub{c} \ce{AA} length and \fsub{d} occupied volume fraction $\focc$. The bead radius is $\SI{1}{\nano\meter}$, and the crowder radius is $\Rc=\SI{5.1}{\nano\meter}$, corresponding to Ficoll 70. For the cytoplasm, $\focc\approx\SI{42.6}{\percent}$.
    }
\label{fig:allosteric:sim}
\end{figure}

\paragraph{Polymer bead models of divalent molecules.}
Polymer bead chains are simplistic models of macromolecules in which ring-forming end-to-end reactions may represent the folding of protein or single-stranded DNA or other conformational changes \cite{ChengThesis}. 
We consider a cyclic \ce{AA} molecule of $n_{\ce{AA}}$ beads that opens to bind a first \ce{B} molecule and closes again after binding another \ce{B} molecule (\cref{fig:allosteric:sim}a).\sknote{bio examples?} 
\Cref{fig:allosteric:sim}b shows the gyration radii of the \ce{AA}, \ce{AA*B}, and \ce{AA*B_2} molecule complexes. 
For $n_{\ce{AA}}>2$, an \ce{AA} first swells and then shrinks upon binding two \ce{B} molecules
\footnote[2]{The gyration radius of the linear chain agrees well with that of a self-avoiding random walk on a BCC lattice in 3D (\cref{sm:fig:Rg}).}.
This case, therefore, corresponds to point 2 of \cref{fig:allosteric:Minton}b, for which SPT predicted enhanced cooperativity. 
Simulation results shown in \cref{fig:allosteric:sim}c,d confirm that the cooperativity increases with crowding for all polymer lengths. 
The magnitude of the observed crowding-induced cooperativity change increases with increasing \ce{AA} length, likely because the differences in the gyration radii increase with the number of beads. 
We also studied linear \ce{AA} polymers and \ce{AA} polymers binding to divalent \ce{BB} particles, and again found increasing cooperativity with increasing \ce{AA} length (\scref{sm:fig:sim:pol1,sm:fig:sim:pol3}).

Unlike the models we have discussed so far, the cytoplasm contains crowders of various shapes and sizes. 
To account for polydispersity in size, we use a mixture of spherical crowders of different radii modeled after the cytoplasm of \emph{E. coli} \cite{Ando2010}.
For this cytoplasm model, cooperativity of divalent binding again increases with the \ce{AA} length (\cref{fig:allosteric:sim}c). 
For the largest \ce{AA} considered (30 beads), the reaction is more than seven times more cooperative inside the cytoplasm than in infinite dilution.
Note a crossover between the cooperativities in the cytoplasm ($\focc\approx \SI{42.6}{\percent}$) and the most crowded Ficoll 70 system ($\focc\approx\SI{44.2}{\percent}$). 
A similar crossover has been reported in the case of macromolecular diffusion \cite{Grimaldo2019}, underlining the importance of the composition of a crowded environment \cite{Kondrat2015, FrembgenKesner2009, Miyaguchi2020}.

\begin{figure}[]
	\includegraphics[width=0.9\linewidth]{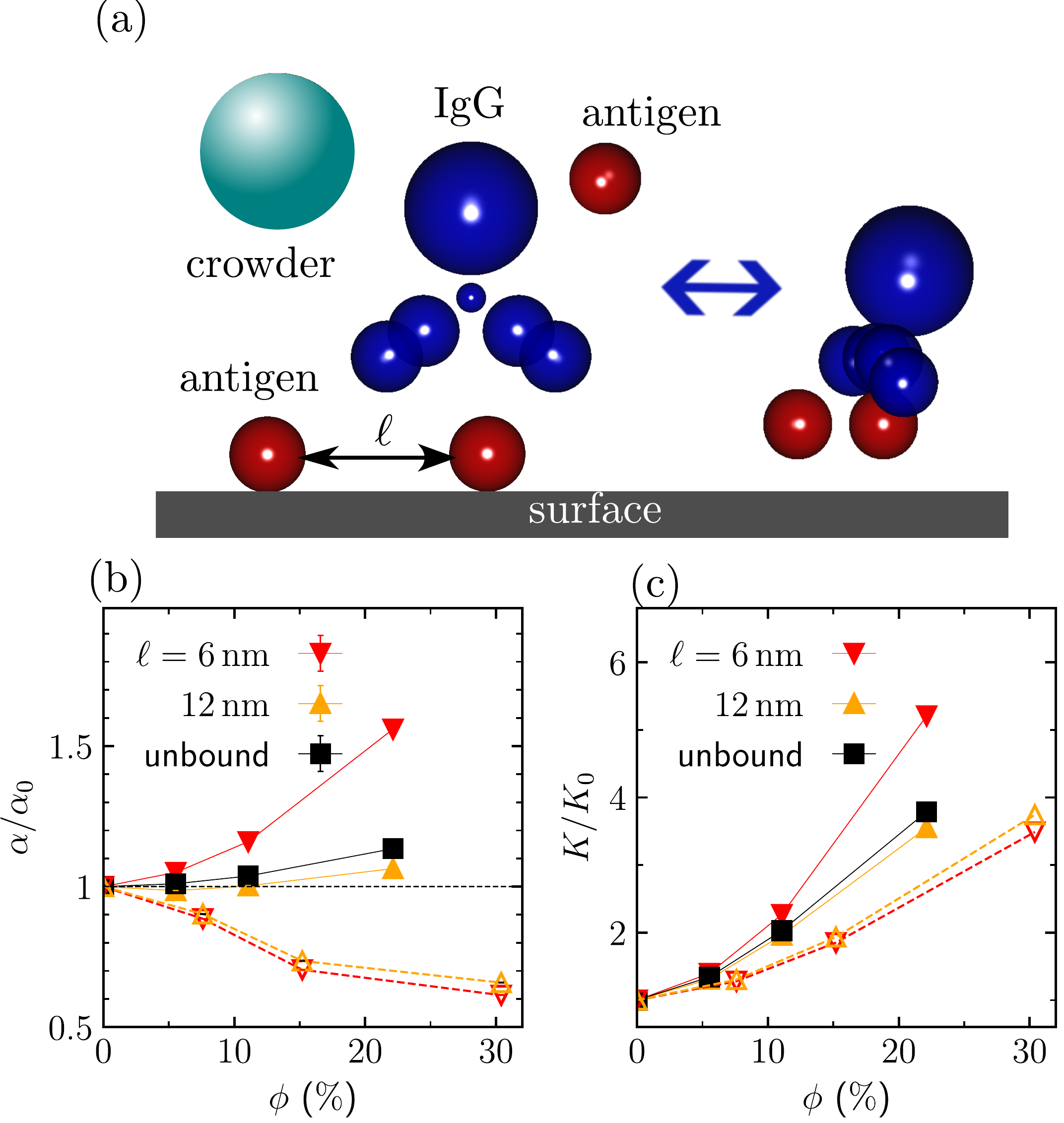}
    \centering
    \caption{
	    \ftitle{IgG--antigen binding in crowded environments.}
	    \fsub{a} Schematics showing a model IgG, antigens, and a crowder. An IgG binds to antigens by two flexible arms modelled by two beads each. 
	    We consider binding in bulk and at a surface. 
	\fsub{b} Cooperativity for IgG-antigen binding as a function of the occupied volume fraction of crowders $\focc$. The results for binding in bulk and on a surface are shown by full and open symbols with the same color code, respectively.
	    \fsub{c} The same as (b) but for the total equilibrium constant.
	}
\label{fig:IgG}
\end{figure}

\paragraph{Immunoglobulin G (IgG)--antigen binding.}
After these generic considerations, we focus on a biologically-relevant example of divalent binding of IgG, which is one of the most common types of antibody.\phantomsection{\label{ref:igg}}~
IgG is a Y-shaped protein, each tip of which binds to the antigen of a microbe or infected cell.
To determine how crowders affect IgG-antigen binding, we use a coarse-grained IgG model, which we introduce in an upcoming article \cite{Slyk2022b} (see \ssect{sec:IgGmodel}). 
This model reproduces the experimentally measured angle distributions \cite{Bongini2004} and hydrodynamic radius of IgG \cite{Junker2019}.
Since IgG binds to antigens on the surface of a pathogen, we consider divalent binding at a flat surface.
For comparison, however, we first evaluate the cooperativity of IgG-antigen binding in bulk, both for unconstrained antigens and when two antigens are at different fixed separations $\ell$ (the latter yields a system with chelate cooperativity).
In all cases, the cooperativity increases with crowding (\cref{fig:IgG}b, full triangles), similar to the polymer bead model considered above. 
The cooperativity increases most for small $\ell$, likely because the crowders squeeze an IgG, making it more likely to bind antigens connected at smaller separations. 

Unlike in bulk, the IgG-antigen cooperativity decreases when the binding occurs at a surface (\cref{fig:IgG}b, open triangles).
Arguably, crowding promotes the binding of a first antigen because an IgG-antigen complex at the surface excludes less volume than a similar complex in bulk;
the same mechanism does not promote the binding of a second antigen, as the partially-bound IgG is already at the surface (\scref{sm:fig:sim:IgG_Ki}).
The equilibrium constant $K_1$ thus increases with crowding more than $K_2$, and hence $\alpha =K_2/K_1$ decreases. 
Still, we observe an increase in the total equilibrium constant $K=K_1 K_2$ in all cases (\cref{fig:IgG}c), in line with the traditional view that crowding enhances association.

\paragraph{Conclusion.}
We have studied how crowding affects the binding of a divalent molecule \ce{AA} to two monovalent molecules \ce{B}. 
By using SPT, we found that crowding could enhance or reduce cooperativity, depending on the sizes of the different molecular complexes (\cref{fig:allosteric:Minton}).
When the \ce{AA} molecules substantially outnumber the \ce{B} molecules, divalent binding only occurs in the presence of crowders.
We developed a general simulation method to compute activity coefficients under crowded conditions and applied it to bead-chain \ce{AA} molecules binding to single-bead \ce{B} molecules  (\cref{fig:allosteric:sim}), corroborating our SPT findings.
We also studied how crowding affects IgG-antigen binding using coarse-grained models for both molecules.
We found opposite trends, where crowding enhances cooperativity for bulk reactions but reduces it when the antibody binds to antigens on a surface. 
Our results suggest that crowding can substantially affect divalent binding, which we hope stimulates further theoretical and experimental studies.
\label{ref1:kinetics}Future work may focus on applying our simulation method to atomistic models of reacting biomolecules and investigating the kinetics of divalent binding in biologically-relevant crowded environments.

This work was supported by NCN grant No. 2017/25/B/ST3/02456 to S.K and T.S. 
We thank PLGrid for providing computational resources.
M.J. was supported by an Advanced Grant from the European Research Council (no. 788954).
The research leading to these results has received funding from the European Union’s Horizon 2020 research and innovation programme under the Marie Skłodowska-Curie grant agreement No 801133.

\bibliography{diffusion-crowding,other}
\end{document}